# A CROSS-VENDOR AND CROSS-STATE ANALYSIS OF THE GPS-PROBE DATA LATENCY

**Zhongxiang Wang**
Ph.D. Student
Department of Civil & Environmental Engineering
University of Maryland
Email: zxwang25@umd.edu

**Masoud Hamedi***
Senior Research Scientist
Center for Advanced Transportation Technology
University of Maryland
5000 College Ave
College Park, MD 20740
Email: masoud@umd.edu | Phone:301-405-2350

**Elham Sharifi**
Faculty Research Assistant
Center for Advanced Transportation Technology
University of Maryland
Email: esharifi@umd.edu

**Stanley Young**
Advanced Transportation and Urban Scientist
National Renewable Energy Laboratory
Email: stanley.young@nrel.gov

Word count: 3,553 words text + (5 tables +8 figures) x 250 words (each) = 6,803 words

Submission date: August 1, 2017

# ABSTRACT

Crowdsourced GPS probe data has become a major source of real-time traffic information applications. In addition to traditional traveler advisory systems such as dynamic message signs (DMS) and 511 systems, probe data is being used for automatic incident detection, Integrated Corridor Management (ICM), end of queue warning systems, and mobility-related smartphone applications. Several private sector vendors offer minute by minute network-wide travel time and speed probe data. The quality of such data in terms of deviation of the reported travel time and speeds from ground-truth has been extensively studied in recent years, and as a result concerns over the accuracy of probe data has mostly faded away. However, the latency of probe data, defined as the lag between the time that disturbance in traffic speed is reported in the outsourced data feed, and the time that the traffic is perturbed, has become a subject of interest. The extent of latency of probe data for real-time applications is critical, so it is important to have a good understanding of the amount of latency and its influencing factors. This paper uses high-quality independent Bluetooth/Wi-Fi re-identification data collected on multiple freeway segments in three different states, to measure the latency of the vehicle probe data provided by three major vendors. The statistical distribution of the latency and its sensitivity to speed slowdown and recovery periods are discussed.
.

# INTRODUCTION

In the era of Big-Data and Internet of Things (IoT), the proliferation of connected devices that continuously report their location has made it possible to capture movement pattern of such devices. Private companies capture the location and speed samples of such devices on the transportation network, and through propriety algorithms and processes, convert them into real-time traffic data products. As a result, minute by minute travel time and speed data over the entire freeway and major arterial road network are available. Probe data has been gaining popularity in recent years, and has quickly become a vital component of Intelligent Transportation Systems. In addition to traditional traveler advisory systems such as dynamic message signs (DMS) and 511 systems, probe data is being used for automatic incident detection, Integrated Corridor Management (ICM), end of queue warning systems, and mobility related smart phone applications. The quality of such data in terms of deviation of the reported travel time and speeds from ground-truth has been extensively studied by independent researchers in recent years. The I-95 Corridor Coalition's Vehicle Probe Project (VPP) data validation program is an example, in which quality of probe data from three major vendors in participating states has been carefully analyzed periodically (http://i95coalition.org/projects/vehicle-probe-project/). As a result, concerns over the accuracy of probe data has mostly faded away. However, the latency of probe data - defined as the lag between the time that a disturbance in traffic speed is reported in the outsourced data feed, and the time that the traffic is perturbed – has become a subject of interest. Latency is inherent in the process of probe data generation due to the processing, communications, and aggregations functions. Although it can be minimized, theoretically speaking it cannot be eliminated. So for real-time applications it is important to have a good understanding of the magnitude latency and its influencing factors.

Latency measurement for real-time travel time data is not well studied. Liu et al. (1) describe the existence of latency in reporting GPS-based data. Tseng et al. (2) discuss a method for traffic sensing with GPS equipped probe vehicles and point out the latency as an outstanding issue. Hunter (3) developed a low-latency state estimation algorithm trying to achieve near real-time measurement of travel time and speed using GPS trajectories on a large-scale network. Chase et al. (4) further stated that GPS-probe data has greater latency when travel speed recovers after peak period, than the beginning of a peak period. None of these papers proposed a detailed quantitative latency measurement methodology.

Kim and Coifman (5) measured the latency for GPS-probe data compared to loop detector. They calculated the correlation coefficient, which significantly depends on the covariance of original time-series speed data and shifted time-series speed data. The results show that the average latency for GPS-probe data is 6.8 minutes, and it could exceed 10 minutes in many situations. In a study by Iowa Department of Transportation (6), and using stationary sensor data, latency for a probe data feed was estimated to vary from 3 to 12 minutes. However, loop detector can only report spot travel speed whereas GPS probe data is reported on standardized segments, typically Traffic Message Channel (TMC) industry standard segments. The speed reported is more related to the space mean speed across a TMC segment rather than a spot speed from loop detectors. Moreover, Kim and Coifman (5) used 10 second aggregation intervals, while granularity of commercially available GPS probe data is one minute or more.

Authors of this paper proposed and published a methodology for calculating latency of GPS probe data compared to a secondary independent data source (Wang et al., (7)). The core of the methodology is a maximum pattern matching algorithm that measures the latency by finding the time shift that maximizes the overlapping of two data sets based on different objective functions. Although the methodology was proven to be effective in explaining both the significance and variation of the latency, it was only applied to a limited data set. This paper employs the same approach, and uses a rich data set comprising of high quality independent Bluetooth/WiFi re-identification data collected on multiple freeway segments in three different states, to measure latency of the vehicle probe data provided by three major vendors. Statistical distribution of the latency and its sensitivity to speed slowdown and recovery periods are also discussed.

## METHODOLOGY
This paper uses the definition and implements the methodology by Wang et al. (7) to measure and analyze the latency of probe data. It includes data preparation, data filtering, interpolation, smoothing and finally data shifting to calculate latency.

### Bluetooth and Wi-Fi Data Fusion
The reference data in this paper is the fused Bluetooth/Wi-Fi re-identification data collected using portable sensors in three states. The basic idea is that two sensors are deployed at the start and end boundaries of a road segment, recording the time stamp and unique digital identifier (MACID) of in-vehicle electronic devices. By matching MACID's between two sensors, a travel time sample is generated. Details of the methodology for matching, filtering and data aggregation can be found in Wang et al. (7). In total, data were collected on 41 segments for approximately 40 days. Table 1 is a summary of the data collection effort, and figure 1 shows the location of the sensors and segments on the map.

TABLE 1. Freeway corridors and Bluetooth/Wi-Fi data collection summary

| State | Road | Start Point | End Point | Length (mile) | Date |
|---|---|---|---|---|---|
| South Carolina | I-85 | US-276/Exit 48 | SC-14/Exit 56 | 7.15 | 12/03/2015 - 12/17/2015 |
| | I-26 | Bush River Rd/Exit 108 | Harbison Blvd/Exit 103 | 4.47 | |
| New Hampshire | I-89 | I-93 | Stickney Hill Rd/Exit 3 | 3.54 | 07/10/2016 - 07/24/2016 |
| | I-93 | I-393/US-202/US-4/Exit 15 | Hackett Hill Rd/Exit 11 | 15.82 | |
| | I-93 | NH-28/Rockingham Rd/Exit 5 | NH-102/Nashua Rd/Exit 4 | 3.63 | |
| North Carolina | I-240 | US-70/Charlotte St/Exit 5B | US-23/US-19/Exit 3 | 2.23 | 12/16/2016 - 12/26/2016 |
| | I-40 | NC-191/Exit 47 | US-23/US-19/Exit 44 | 2.56 | |
| | I-26 | I-40/ Exit 46 A/US 74 | I-26/Exit 37 | 14.43 | |

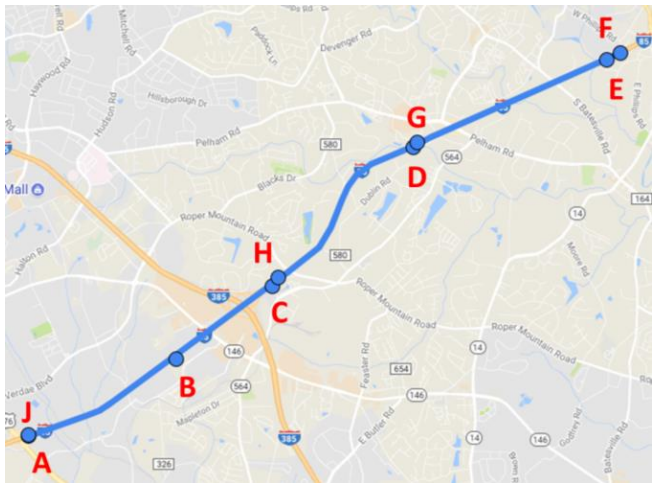

(a) SC (I-85)

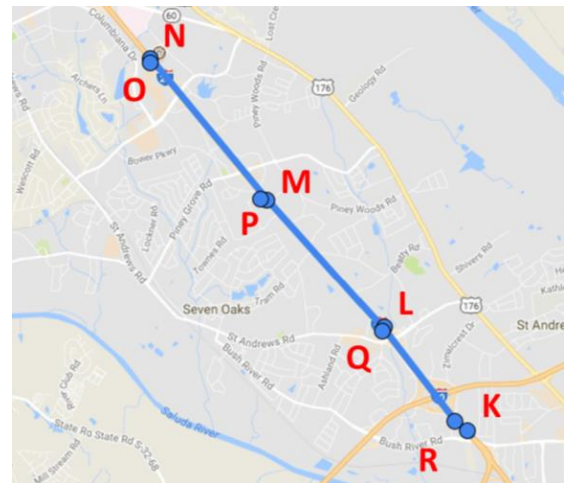

(b) SC (I-26)

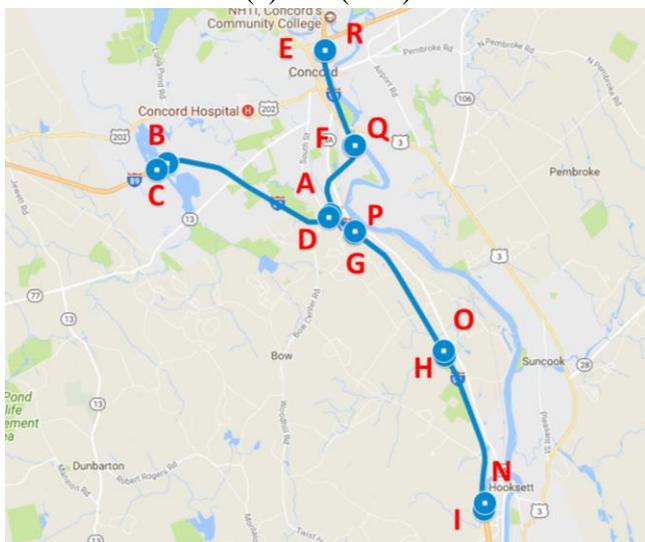

(c) NH (I-89, I-93)

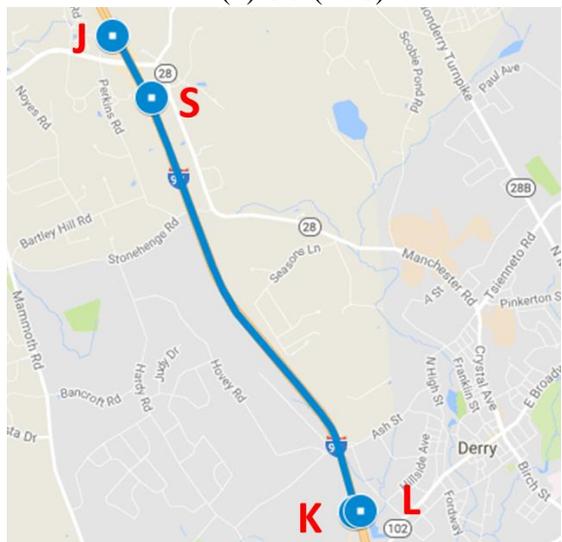

(d) NH (I-93)

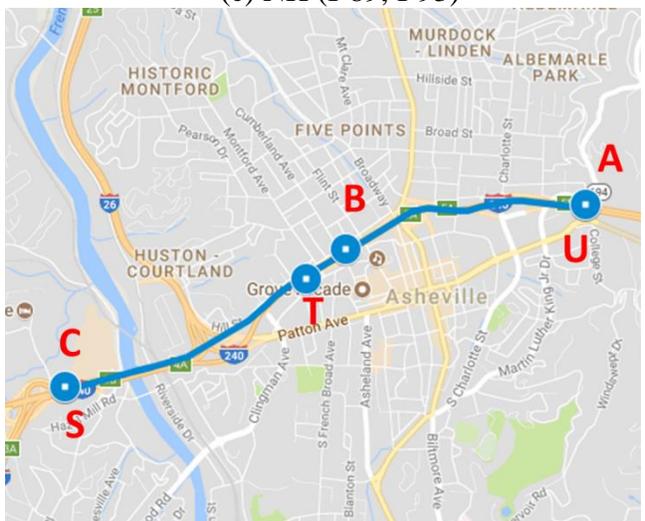

(e) NC (I-240)

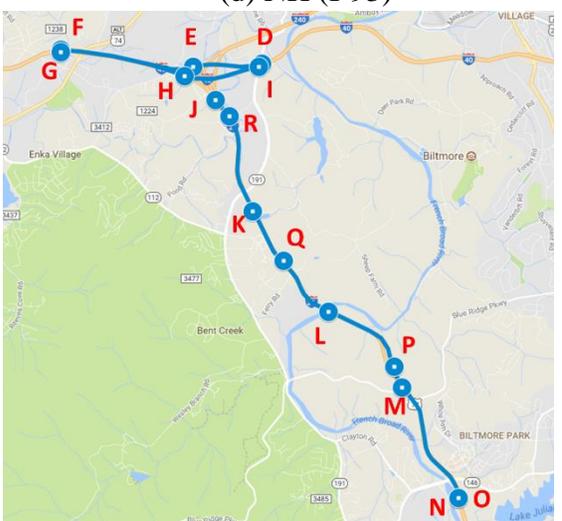

(f) NC (I-40, I-26)

FIGURE 1.  Freeway test segments and Bluetooth/Wi-Fi sensor locations

**Capturing Slowdown and Recovery Episodes**

Average travel speed on freeways is subject to fluctuation. Recurring congestion is caused by increased traffic volume and insufficient capacity, while non-recurring congestion often results from incidents, work zones, weather events and special events that impact traffic flow. When traffic speeds drop significantly for a considerable amount of time, a slowdown episode emerges. In such occasions, traffic speeds will eventually bounce back to the pre-congestion levels. Figure 2 shows an example of slowdown and recovery, captured by both Bluetooth and GPS probe data. In order to measure the latency of probe data with respect to the reference data, start and end time of each Slowdown and Recovery Episode (SRE) as well as the transition time between Slowdown Period (SP) to Recovery Period (RP) in the collected data must be identified. In the previous effort to measure the latency of probe data (7), authors visually inspected the daily speed graphs to record slowdowns. However, manual approach is tedious and analyzing a large number of segments for an extended time period calls for automation.

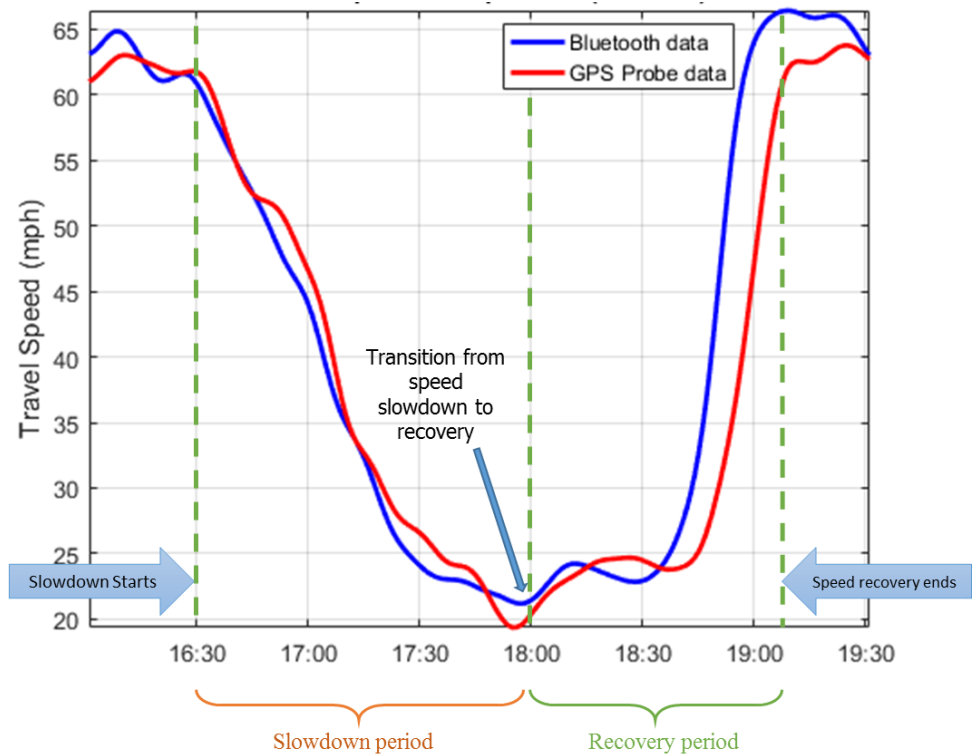

FIGURE 2. Illustration of traffic slowdown and recovery

A pattern recognition algorithm was developed. The main steps of Automated Speed Pattern Recognition Filter (ASPRF) are as follows:

For every path in the data set:

  For every day of the study period:

   • Calculate free flow speed (FFS) as the 80th percentile of travel speed, for each specific path at each day and the minimum slowdown threshold (MST) as the 40% of the free flow speed;

- Mark all data points with speed less than MST (there could exist several data groups for each day, each group is a potential SRE);
  - For each group:
    - Find the minimum speed point and its corresponding time is the transition from SP to RP;
    - From the minimum speed point, search both forward and backward until the speed reaches the FFS, the first reaching time is the start time of the SP (for backward search) and the end time of the RP (for forward search); if FFS cannot be reached, switch to alternative boundaries for SRE (start and end time of each day, the end of previous RP and the start of the succeeding SP);
- Rule-based SRE filtering:
  - Discard any SRE with missing speed data for five or more consecutive minutes.
  - Patch any SRE with missing speed data for less than five consecutive minutes, by interpolating neighboring data, as discussed by Wang et al. (7)
  - Combine two adjacent SREs if the end of the first RP is within 30 minutes of the start of the second SP;
  - Discard any SRE if its final duration is less than 60 minutes.

After implementing the ASPRF in Python, it was applied on the reference data set, and the results were visualized and inspected. Figure 3 shows an example of the algorithm output for one day on one path in South Carolina. The red dots mark the start of the SP, the transition from SP to RP and the end of the RP. The upper dashed orange line is the FFS and the lower dash-dotted orange line is MST. Blue and green shaded areas show SP and RP respectively. Although there is a sudden speed in the early hours of the day, it is not detected as a slowdown since for latency measurement only significant slowdowns with at least one-hour duration are considered.

In order to verify the ASPRF output and calibrate the thresholds, it was applied to the same data set used in (7) and results were found to be compatible with the manual process. Table 2 shows the number of slowdown and recovery episodes for each state after running the code.

TABLE 2. Number of slowdown and recovery episodes per state

| State | # of paths | # of days | # of Slowdown and Recovery Episodes |
|-------|------------|-----------|-------------------------------------|
| SC | 14 | 15 | 72 |
| NH | 12 | 15 | 26 |
| NC | 15 | 11 | 18 |
| **Total** | **41** | **41** | **116** |

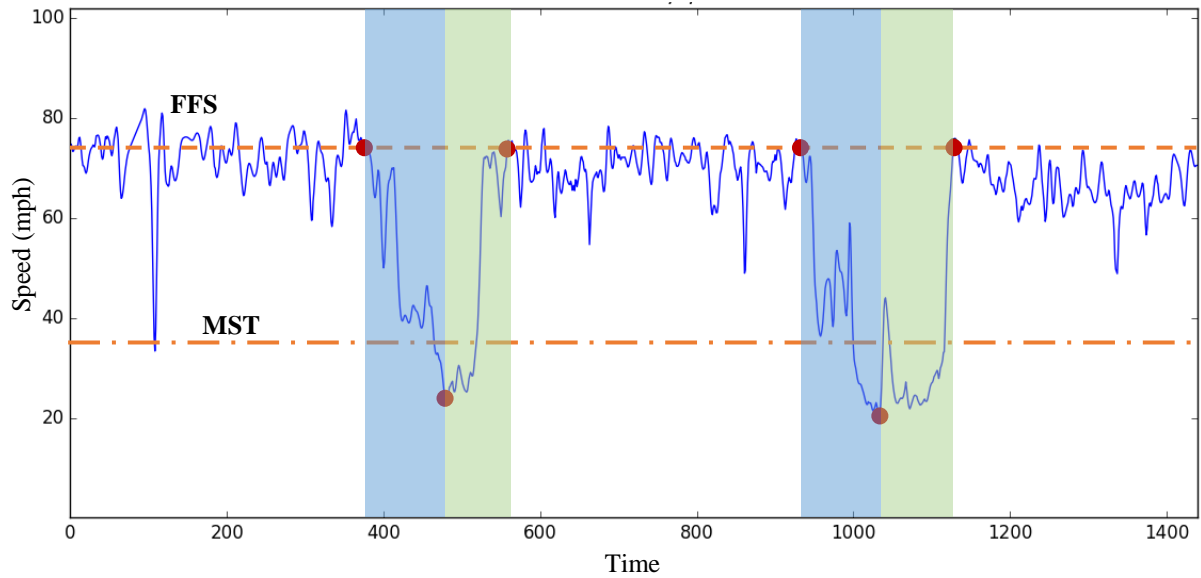

FIGURE 3. Sample slowdown and recovery detection results from ASPRF algorithm

**Latency Measurement**

As mentioned before, the same methodology by Wang et al. (7) is used in this paper. The basic idea is to shift the GPS-probe data from a lower bound to an upper bound minute by minute, and find the time shift that yields maximum overlap between the probe and the reference data. Three different fitness functions are used to measure the overlap:

1. **AVD**: minimize sum of absolute speed different between GPS-probe data and reference data for all corresponding minute by minute data points;
2. **SVD**: minimize sum of squared speed difference between GPS-probe data and reference data for all corresponding minute by minute data points;
3. **COR**: maximize correlation coefficient between GPS-probe data and reference data.

Each of the above fitness functions provides a different perspective for the curve matching. AVD is the most straightforward way to measure and minimize vertical distance between the two curves. SVD puts more emphasis on higher speed differences. COR describes the statistical correlation between two curves. As discussed in (7), by applying all three fitness functions separately, and measuring the latency based on their average, the methodology becomes less sensitive to the irregularities in data and produces more consistent results.

## TEST RESULTS

The reference data was collected on selected freeway corridors in South Carolina, New Hampshire, and North Carolina as discussed in Table 1. Corridors were divided into shorter paths, with length varying from 1 to 3.5 miles. Approximately two-week worth of data was collected in each state. GPS probe data from three major vendors was obtained from the I-95 Corridor Coalition as part of their Vehicle Probe Project. The purpose of this study is to investigate the latency of GPS probe data in general, and not to scrutinize data vendors. So the vendors will be referred by numbers when discussing the results.

The latency measurement procedure was applied to data for all three vendors, and for all 116 SRE, SD and RP cases identified by ASPRF algorithm. Figure 4 shows the results for segment F to G in South Carolina on 12/10/2015 as a sample.

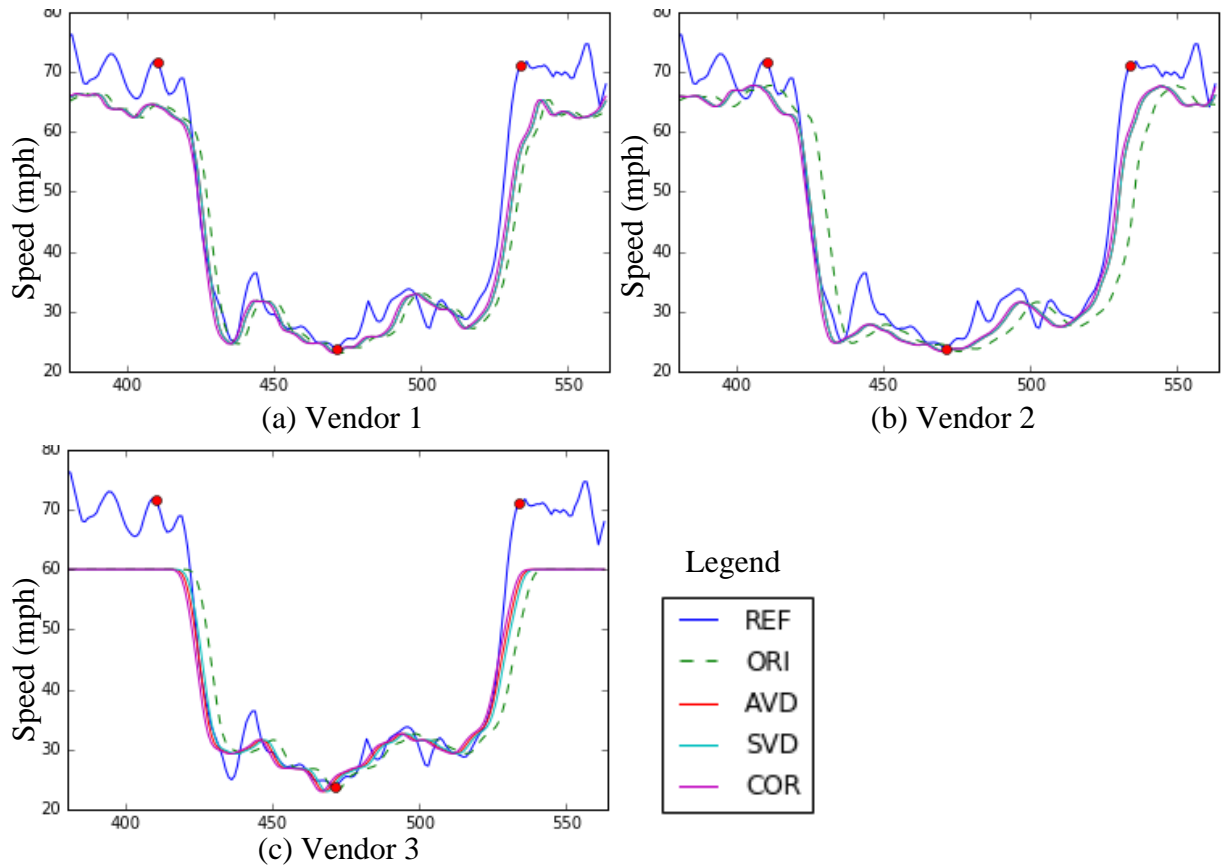

(a) Vendor 1

(b) Vendor 2

(c) Vendor 3

Legend

Note: REF: reference data; ORI: original un-shifted GPS-probe data; AVD: GPS-probe data shifted and latency measured by AVD; SVD: GPS-probe data shifted and latency measured by SVD; COR: GPS-probe data shifted and latency measured by COR

**FIGURE 4. Latency measurement results of segment F-G in South Carolina for 12/10/2015**

Figure 4 shows that all three vendors have successfully captured both speed reduction and recovery in their data feed. For each fitness function, the shifted graph corresponding to the optimal value is shown, and the shift value is reported as measured latency. Also, difference among shift results based on the three applied fitness functions is marginal. The red dots on the graphs show the start time of the slowdown, start time of the recovery and the end time obtained from ASPRF.

After running the latency measurement procedure for all cases and all three vendors, graphs similar to figure 1 generated and carefully examined. Aggregated numerical results are presented in table 3. In each column, average latency values obtained based on the fitness function in the column header are reported. The column marked as "mean" is the statistical mean of the latency across all three fitness functions, and is the value that is used as the bottom line.

**TABLE 1. Aggregated latency measurement results for all vendors**

| | | Vendor 1 | | | | Vendor 2 | | | | Vendor 3 | | | |
|---|---|---|---|---|---|---|---|---|---|---|---|---|---|
| | | Average Latency (minute) | | | | Average Latency (minute) | | | | Average Latency (minute) | | | |
| | | AVD | SQR | COR | **Mean** | AVD | SQR | COR | **Mean** | AVD | SQR | COR | **Mean** |
| **SC** | SRE | 3.9 | 4.4 | 4.6 | **4.3** | 6.1 | 6.2 | 6.2 | **6.2** | 2.9 | 3.1 | 3.1 | **3.0** |
| | SP | 3.6 | 3.6 | 4.0 | **3.7** | 6.7 | 6.8 | 6.6 | **6.7** | 4.4 | 4.3 | 3.8 | **4.2** |
| | RP | 4.7 | 5.3 | 4.0 | **4.7** | 5.0 | 5.2 | 5.0 | **5.1** | 2.0 | 2.4 | 2.5 | **2.3** |
| **NH** | SRE | 3.3 | 3.4 | 3.6 | **3.4** | 5.6 | 6.0 | 6.4 | **6.0** | 2.3 | 2.5 | 2.7 | **2.5** |
| | SP | 3.3 | 3.2 | 3.2 | **3.2** | 7.0 | 7.0 | 6.7 | **6.9** | 3.5 | 3.0 | 3.2 | **3.2** |
| | RP | 4.0 | 4.2 | 3.9 | **4.0** | 3.6 | 3.7 | 4.7 | **4.0** | 2.9 | 3.5 | 2.6 | **3.0** |
| **NC** | SRE | 3.5 | 4.1 | 4.1 | **3.9** | 7.3 | 7.8 | 7.3 | **7.5** | 3.2 | 3.2 | 2.7 | **3.0** |
| | SP | 4.6 | 4.6 | 3.6 | **4.2** | 7.7 | 7.8 | 7.4 | **7.6** | 3.2 | 2.9 | 2.5 | **2.9** |
| | RP | 3.4 | 3.7 | 4.1 | **3.7** | 5.9 | 6.5 | 6.8 | **6.4** | 3.2 | 3.1 | 3.8 | **3.4** |

Note: SC: South Carolina, NH: New Hampshire, NC: North Carolina, AVD: average latency based on absolute vertical distance fitness function; SQR: average latency based on squared vertical distance fitness function; COR: average latency based on correlation coefficient fitness function; SRE: Slowdown and Recovery Episode; SP: Slowdown Period; RP: Recovery Period

Table 3 shows that variation of latency from vendor to vendor. For example, for all test segments in NC, and for the entire slowdown and recovery episodes (SRE), vendor 3 has the smallest average latency value of 2.5 minutes, while this number is 3.4 and 6.0 minutes for vendors 1 and 2 respectively. Looking at results across states for each vendor, shows that the mean SRE latency value ranges from 3.4 to 4.3, 6.0 to 7.5 and 2.5 to 3.0 for vendors 1, 2 and 3 respectively. Vendor 1 has the smallest range suggesting that its latency is more consistent.

In their previous work (7), authors had observed that latency is not symmetric for each SRE, meaning that the probe data feed might reflect the reduction in the speed with less latency compared to the speed recovery. To revisit that observation, all SRE cases are examined and the percentage of cases that show higher latency in SP compared to RP are reported in table 4.

**TABLE 2. Percentage of slowdown and recovery episodes with slowdown latency greater than recovery latency**

| State | Vendor 1 | Vendor 2 | Vendor 3 |
|---|---|---|---|
| SC | 38% | 64% | 68% |
| NH | 50% | 77% | 50% |
| NC | 50% | 72% | 56% |
| Total | 42% | 68% | 62% |

As Table 4 shows, the results vary by the vendor and state. So although there is asymmetry in the latency between slowdown and recovery periods, it cannot be said that in general SP latency is higher than RP latency or vice versa. Having said that, the last row of Table 4 shows that cases with SP latency greater than RP latency are in minority (42%) for vendor 1 compared to the other two vendors.

**Latency distribution**
Results in Tables 3 and 4 show that the probe data latency varies from vendor to vendor, and from location to location. So instead of reducing latency to a single value, it is important to study its statistical distribution. Figure 5 shows box plot of the latency observations created for all latency observation during the entire slowdown and recovery episode, slowdown period and recovery periods separated by the vendor. The box plots show that latency observations for the vendor 2 are more scattered compared to the other two vendors. Recovery periods latency for vendor 2 is also more evenly distributed compared to slowdown period for the same vendor. Same results are illustrated in figure 6 from a different perspective as histogram plots.

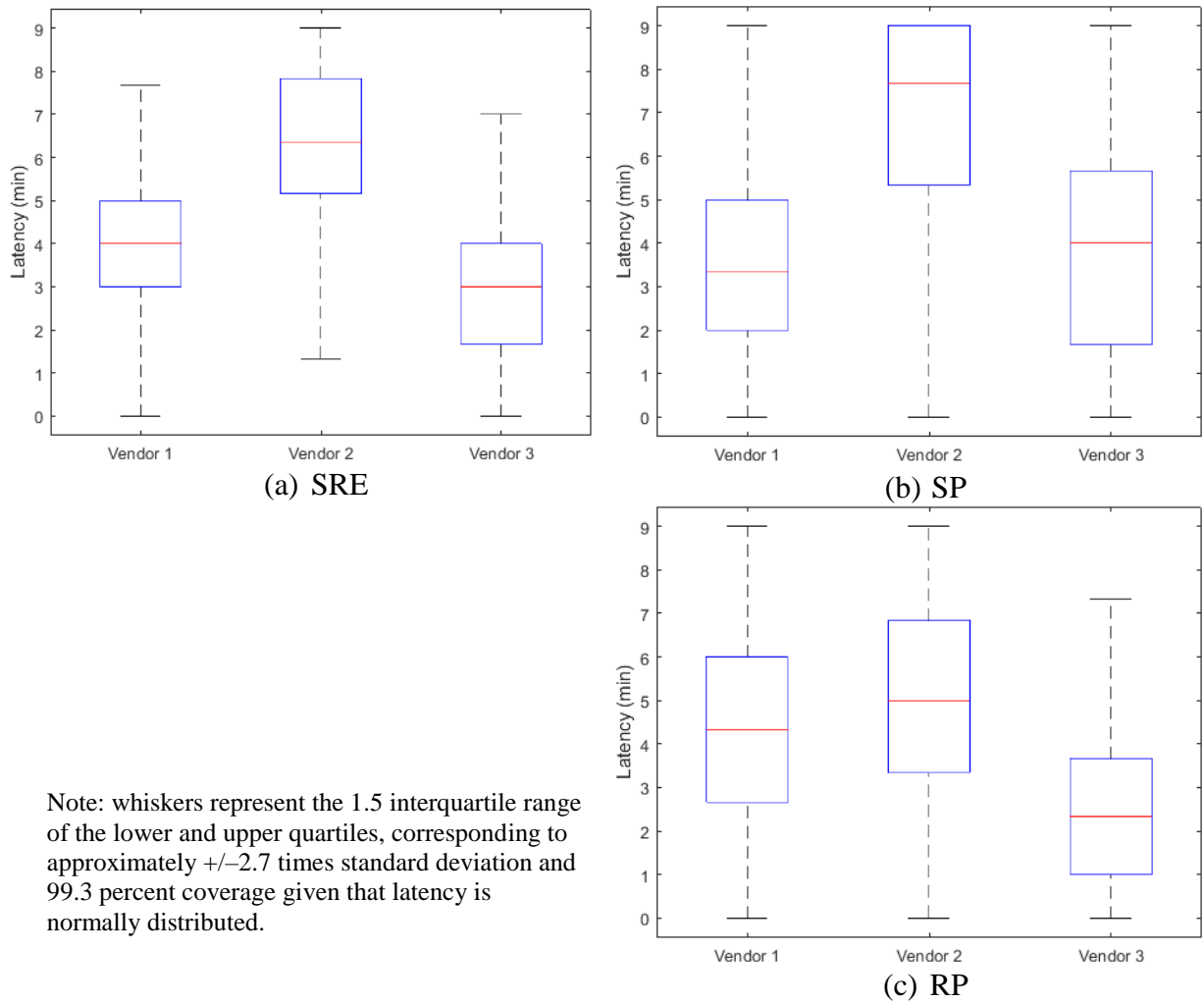

Note: whiskers represent the 1.5 interquartile range of the lower and upper quartiles, corresponding to approximately +/–2.7 times standard deviation and 99.3 percent coverage given that latency is normally distributed.

**FIGURE 1  Latency box plot (a) entire Slowdown and Recovery Episode (b) Slowdown Period (c) Recovery Period**

**Statistical comparing among vendors**
As shown in Figures 5 and 6, statistical distribution for latency observations is available for each vendor. So it is possible to examine whether mean latency reported by a vendor is statistically equal to the mean of other vendors. A T-test is used to test such hypothesis:

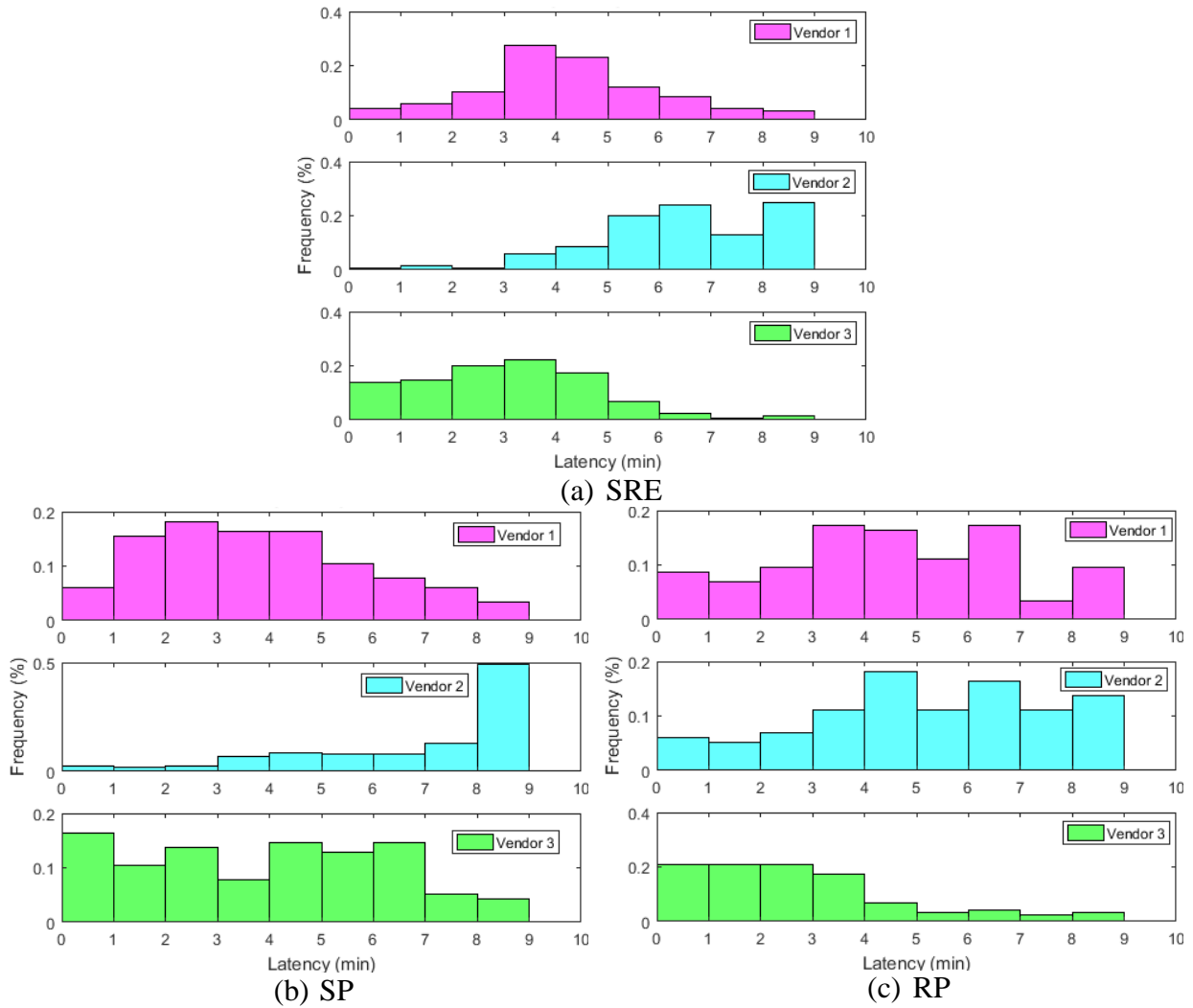

**FIGURE 6. Latency histogram (a) entire Slowdown and Recovery Episode (b) Slowdown Period (c) Recovery Period**

$H0 : \mu_a - \mu_b = 0$

$H1 : \mu_a - \mu_b \neq 0$    where $\mu_a$ is the average latency of vendor a, and $\mu_{b1}$ is the average latency of vendor b

T-test results for each pair of vendors are shown in table 5, where the numbers above diagonal line are t-statistics, and the numbers below the line are P-values for 99% confidence level. The results show that average latency between each pair of vendors is statistically different, so there is difference among the vendors when it comes to the latency.

**TABLE 5. T-test pairwise comparison of latency**

| P-value | t-stat | Vendor 1 | Vendor 2 | Vendor 3 |
|---|---|---|---|---|
| Vendor 1 | | | **13.09** | **6.02** |
| Vendor 2 | | 0.00 | | **20.21** |
| Vendor 3 | | 0.00 | 0.00 | |

**Sensitivity of latency to the time of the day**

Results in (7) suggested that there was not a meaningful difference between latency in the morning compared to the afternoon hours. To revisit that result, latency observations for morning and afternoon travel in this paper were separated and analyzed. Figure 7 shows the results for each vendor and for SRE, SP and RP. This figure confirms the previous conclusion that there is no consistent pattern in the results across different cases and vendors. However, one interesting observation is that for SRE, vendor 3 seems to have more observations with high latency values during afternoon hours compared to morning hours.

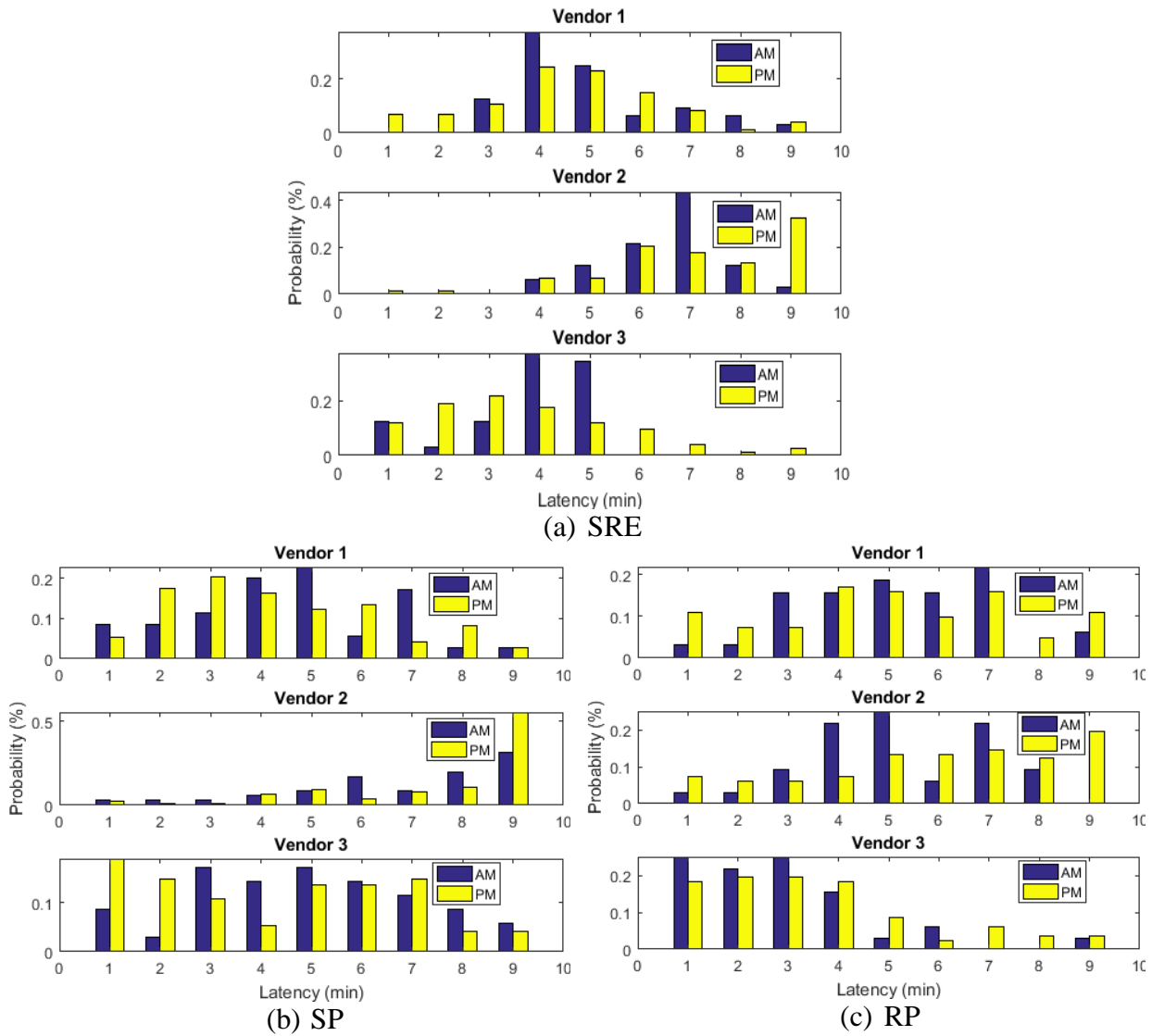

**FIGURE 7. AM vs PM Latency distribution (a) entire Slowdown and Recovery Episode (b) Slowdown Period (c) Recovery Period**

**Sensitivity of latency to the segment length**

Correlation between latency and path length is another plausible hypothesis. Results in (7) show that such correlation does not exist. Results of the latency are presented as a scatter plot in figure 8, where the horizontal axis shows the segment length, and the vertical axis is the latency. As illustrated, there is no strong relationship between length and latency, and the observation yields for all vendors. Statistical examination of the samples confirms this conclusion.

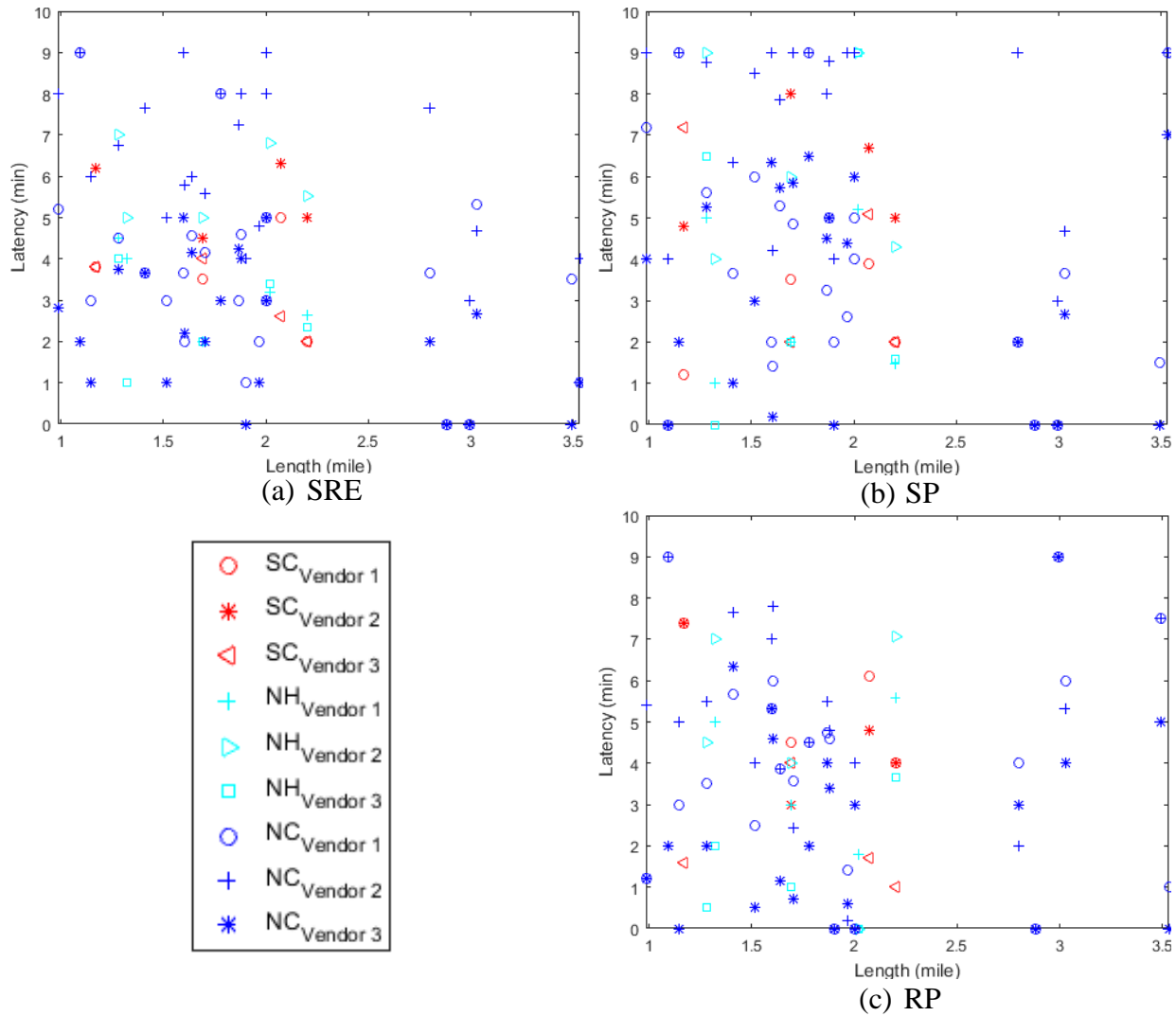

(a) SRE  (b) SP  (c) RP

**FIGURE 8. Latency box plot (a) entire Slowdown and Recovery Episode (b) Slowdown Period (c) Recovery Period**

## CONCLUSIONS

This paper made an effort to measure and report latency of the GPS probe data, with respect to Bluetooth/WiFi re-identification data using a maximum pattern matching algorithm. Reference data was collected on a number of freeway segments in different states and different times of the year. Probe data obtained from three major private sector vendors were examined. Results were presented for the entire slowdown and recovery episodes, and for slowdown and recovery periods separately. Across all vendors on the test roads used in this research, the average latency was found to be 4.4 minute for freeways (signalized arterials were not tested) with a standard deviation in latency of 2.3 minute. Statistically speaking, average latency for different vendors are different. Results show that instead of interpreting the latency as a single number, distribution of latency should be measured and evaluated. No strong correlation between latency and time of the day, and also latency and segment length were found.

Automation of all steps in the solution approach makes it possible to apply the latency measurement procedure to future data sets. Authors plan to test more data sets as they become available, and study possible enhancements in latency over time, as both sampling rate of probe vehicles and performance of data fusion engines used by vendors will improve. Latency of probe data on signalized arterials where speeds are subject to higher fluctuations remains unknown. Changing the methodology in order to accommodate arterial data is subject of future research. Probe data is used in many real-time applications. Understanding implications of the latency and its distribution on such applications and exploring solutions to compensate for latency is another area for future research.

## ACKNOWLEDGEMENTS
Data used in this study was collected by the I-95 Corridor Coalition as part of their Vehicle Probe Project. The results and conclusions in this document are those of the authors and not the I-95 Corridor Coalition.

## REFERENCE
1   Liu, X, Chien, S., and Kim K. "Evaluation of floating car technologies for travel time estimation." Journal of Modern Transportation 20.1 (2012): 49-56.
2   Peng-Jui Tseng, C. C. Hung, Tsung-Hsun Chang and Yu-Hsiang Chuang, "Real-time urban traffic sensing with GPS equipped Probe Vehicles," 2012 12th International Conference on ITS Telecommunications, Taipei, 2012, pp. 306-310.
3   Hunter, Timothy J, Large-Scale, Low-Latency State Estimation of Cyber- Physical Systems With An Application To Traffic Estimation, PhD Dissertation, University of California, Berkeley, 2014
4   Chase, R., et al. "Comparative evaluation of reported speeds from corresponding fixed-point and probe-based detection systems." Transportation Research Record: Journal of the Transportation Research Board 2308 (2012): 110-119.
5   Kim, S., and Coifman, B. Comparing INRIX speed data against concurrent loop detector stations over several months. Transportation Research Part C: Emerging Technologies 49 (2014): 59-72.
6   Center for Transportation Research and Education. (2015). "Evaluation of INRIX Probe Data". August 2015. Sponsored by Iowa Department of Transportation. (Task 10 of InTrans Project 13-480). http://reactor.ctre.iastate.edu/publications/task10.pdf
7   Wang, Z., Hamedi, M. and Young, S., 2017. A Methodology for Calculating Latency of GPS Probe Data. Transportation Research Record: Journal of the Transportation Research Board, No. 2645. DOI: 10.3141/2645-09.